\documentclass[11pt]{article}
\usepackage[a4paper,margin=1in]{geometry}
\usepackage[titletoc,title]{appendix}
\usepackage{changepage}
\usepackage[utf8x]{inputenc}
\usepackage{textcomp,marvosym}
\usepackage{fixltx2e}
\usepackage{amsmath,amssymb}
\usepackage{cite}
\usepackage{nameref,hyperref}
\usepackage{authblk}
\usepackage{algorithm}
\usepackage{algpseudocode}
\usepackage{dsfont}
\usepackage{xcolor}
\usepackage{enumitem}
\usepackage{caption,setspace}
\usepackage{subcaption}
\usepackage{pdflscape}
\usepackage{graphicx}
\usepackage{placeins}
\usepackage{mathtools}

\renewcommand{\eqref}[1]{Equation~(\ref{#1})}

\title{Exact sharp-fronted travelling wave solutions of the Fisher-KPP equation}

\author[1]{Scott~W. McCue}
\author[1]{Maud El-Hachem}
\author[1]{Matthew~J. Simpson\footnote{To whom correspondence should be addressed. E-mail: matthew.simpson@qut.edu.au}}

\affil[1]{School of Mathematical Sciences, Queensland University of Technology, Brisbane, Queensland 4001, Australia}

\begin{document}

\maketitle
\begin{abstract}
A family of travelling wave solutions to the Fisher-KPP equation with speeds $c=\pm 5/\sqrt{6}$ can be expressed exactly using Weierstra$\ss$ elliptic functions.  The well-known solution for $c=5/\sqrt{6}$, which decays to zero in the far-field, is exceptional in the sense that it can be written simply in terms of an exponential function.  This solution has the property that the phase-plane trajectory is a heteroclinic orbit beginning at a saddle point and ends at the origin.  For $c=-5/\sqrt{6}$, there is also a trajectory that begins at the saddle point, but this solution is normally disregarded as being unphysical as it blows up for finite $z$.  We reinterpret this special trajectory as an exact sharp-fronted travelling solution to a \textit{Fisher-Stefan} type moving boundary problem, where the population is receding from, instead of advancing into, an empty space.  By simulating the full moving boundary problem numerically, we demonstrate how time-dependent solutions evolve to this exact travelling solution for large time.  The relevance of such receding travelling waves to mathematical models for cell migration and cell proliferation is also discussed.
\end{abstract}


\newpage
\section{Introduction} \label{sec:intro}

For various applications in ecology and cell biology, the Fisher-KPP equation~\cite{Fisher1937,Kolmogorov1937,Canosa1973}
\begin{equation}
\frac{\partial u}{\partial t}=\frac{\partial^2 u}{\partial x^2}+u(1-u),
\label{eq:FisherKPP}
\end{equation}
provides a very well studied model for the growth and spread of a population of species or cell types~\cite{Murray02,Edelstein2005,Kot03}.  One key mathematical result is that, with the associated initial and boundary conditions
\begin{equation}
u(x,0)=F(x), \quad 0<x<\infty,
\label{eq:FisherKPP_IC}
\end{equation}
\begin{equation}
\frac{\partial u}{\partial x}=0 \quad\mbox{on}\quad x=0,
\quad u\rightarrow 0\quad\mbox{as}\quad x\rightarrow\infty,
\label{eq:FisherKPP_BC}
\end{equation}
the time-dependent solution evolves towards a travelling wave profile $U(z)$, where $z=x-ct$, as $t\rightarrow\infty$~\cite{Fisher1937,Kolmogorov1937,Canosa1973}.  A combination of phase-plane analysis and simple asymptotics demonstrates that the travelling wave speed $c$ satisfies $c\geq 2$ and is selected by the far-field behaviour of the initial condition $F(x)$ in (\ref{eq:FisherKPP_IC})~\cite{Murray02,Edelstein2005,Kot03}.

In this work, we are motivated by our recent studies~\cite{Elhachem2019,Elhachem2020a,Elhachem2020b} where we have restricted the Fisher-KPP equation (\ref{eq:FisherKPP}) to hold on the moving domain $0<x<s(t)$, together with a Stefan-type moving boundary condition, to give the so-called {\em Fisher-Stefan} model~\cite{Du2010}
\begin{equation}
\frac{\partial u}{\partial t}=\frac{\partial^2 u}{\partial x^2}+u(1-u),
\quad 0<x<s(t),
\label{eq:StefanFisher1}
\end{equation}
\begin{equation}
\frac{\partial u}{\partial x}=0 \quad\mbox{on}\quad x=0,
\label{eq:StefanFisher2}
\end{equation}
\begin{equation}
u=0, \quad \frac{\mathrm{d}s}{\mathrm{d}t}=-\kappa\frac{\partial u}{\partial x}
\quad\mbox{on}\quad x=s(t).
\label{eq:StefanFisher3}
\end{equation}
Here $\kappa$ is a parameter that relates the leakage of the population at the boundary to the speed of the boundary.  In this context, we and others have provided new interpretations for travelling wave solutions (\ref{eq:FisherKPP}) for $c<2$, including slowly moving fronts that advance with speed $0<c<2$~\cite{Elhachem2019,Du2010,Du2014a,Du2014b,Du2015}, stationary profiles for $c=0$ and receding fronts with speed $c<0$~\cite{Elhachem2020a,Elhachem2020b}.  Travelling wave solutions for $c<2$ are interesting because they are normally disregarded as being unphysical (because they do not satisfy the boundary conditions and/or are not restricted to $0<U<1$ for all $z\in \mathbb{R}$)~\cite{Fisher1937,Kolmogorov1937,Canosa1973,Murray02,Edelstein2005,Kot03}.

\begin{figure}
	\centering
	\includegraphics[width=1\textwidth]{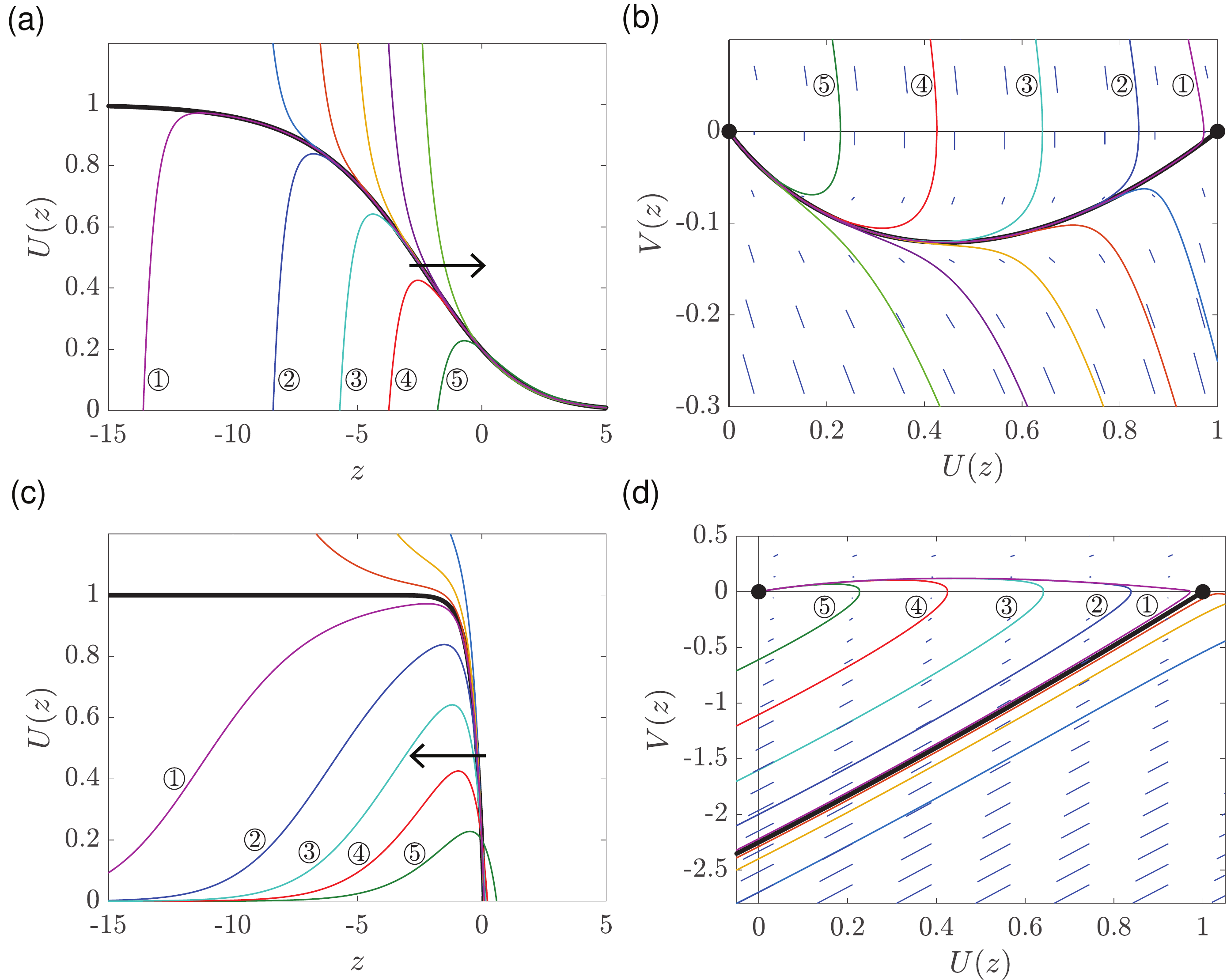}
	\caption{\textbf{Exact solutions to the Fisher-KPP model}. (a)-(b) shows the exact solutions for $c=5/\sqrt{6}$ in the physical plane and the phase plane, respectively. Similarly, (c)-(d) shows the exact solutions for $c=-5/\sqrt{6}$ in the physical plane and the phase plane, respectively. (a), (c) show the travelling wave profiles $U(z)$ that satisfy (\ref{eq:Urightarrow1}) (thick black), horizontal arrows are superimposed to emphasize the difference in direction. (b),(d) show various phase plane trajectories.  Equilibrium points in the phase planes are shown with black discs, and trajectories corresponding to the special travelling wave solutions are plotted (thick black).  Various other trajectories (thin coloured lines) are superimposed, and the corresponding curves are given in (a),(c) using the same colours as in (b),(d). Some trajectories are numbered to emphasise the point that the numbered trajectories in (a)-(d) are identical.}
	\label{fig:figure1}
\end{figure}

In this letter we focus on travelling wave solutions to (\ref{eq:FisherKPP}) for the special values $c=\pm 5/\sqrt{6}$.  For $c=5/\sqrt{6}$ there is a well known exact solution~\cite{Murray02,Kaliappan1984}
\begin{equation}
U=\left(1+(\sqrt{5}-1)\,\mathrm{e}^{z/\sqrt{6}}
\right)^{-2},
\label{eq:exactsoln}
\end{equation}
as shown in Figure \ref{fig:figure1}(a)-(b).  Other exact travelling wave solutions to (\ref{eq:FisherKPP}) for $c=\pm 5/\sqrt{6}$, which can be written in terms of Weierstra$\ss$ elliptic functions~\cite{Ablowitz1979}, are normally disregarded as being unphysical in the usual way~\cite{Murray02}.  However, in the context of the Fisher-Stefan model (\ref{eq:StefanFisher1})-(\ref{eq:StefanFisher3}), we provide a new physical interpretation of one of these solutions. In particular, we claim that one of the profiles for $c=-5/\sqrt{6}$ corresponds to a receding travelling wave to (\ref{eq:StefanFisher1})-(\ref{eq:StefanFisher2}) with a special value of $\kappa=-0.906\ldots$.  In this way, we illustrate a second physically realistic exact travelling wave solution to (\ref{eq:FisherKPP}) for $c=\pm 5/\sqrt{6}$.

In section~\ref{sec:tws} we review the exact travelling solutions to the Fisher-KPP equation for $c=\pm 5/\sqrt{6}$, taken from Ablowitz \& Zeppetella \cite{Ablowitz1979}.  This derivation involves Weierstra$\ss$ elliptic functions~\cite{Abramowitz1965}.  By using phase-plane analysis, we demonstrate the qualitative behaviour of this family of solutions and highlight the two special trajectories that evolve to $(U,V)=(1,0)$ as $z\rightarrow -\infty$.  In section~\ref{sec:FSm} we link this special trajectory to solutions of the Fisher-Stefan model and demonstrate numerically that for $\kappa=0.906\ldots$, initial conditions evolve to this travelling wave solution for large time. Finally, we providing concluding remarks in section~\ref{sec:Conclusion}.

\section{Travelling wave solutions for $c=\pm 5/\sqrt{6}$}\label{sec:tws}

To study travelling wave solutions of (\ref{eq:FisherKPP}) we write $u=U(z)$, where $z=x-ct$, to give
\begin{equation}
\frac{\mathrm{d}^2U}{\mathrm{d}z^2}+c\frac{\mathrm{d}U}{\mathrm{d}z}+U(1-U)=0.
\label{eq:2ndorderode}
\end{equation}
We discuss the domain of interest and the boundary conditions below, but for the moment we highlight the physically relevant boundary condition
\begin{equation}
U\rightarrow 1^-, \quad \frac{\mathrm{d}U}{\mathrm{d}z}\rightarrow 0^-,
\quad\mbox{as}\quad z\rightarrow -\infty,
\label{eq:Urightarrow1}
\end{equation}
which applies for two special cases considered below.  We rewrite (\ref{eq:2ndorderode}) in the usual way as
\begin{align}
	\frac{\text{d}U}{\text{d} z} &= V, \label{eq:1storderodeU}\\
\frac{\text{d}V}{\text{d} z} &= -cV - U(1-U). \label{eq:1storderodeV}
\end{align}
One point to note here is that this system is reversible under the substitution $z\rightarrow -z$, $V\rightarrow -V$, $c\rightarrow -c$, which means that the phase-plane for (\ref{eq:1storderodeU})-(\ref{eq:1storderodeV}) for $c<0$ is simply a reflection about the $U$-axis of the phase-plane for $c>0$.

The concern here is with special values $c=\pm 5/\sqrt{6}$.  For these values, we may solve (\ref{eq:2ndorderode}) exactly, as explained by Ablowitz \& Zeppetella \cite{Ablowitz1979}.  A summary of the working is as follows.  We start by letting $U=f(z)w(z)$ and substituting into (\ref{eq:2ndorderode}).  Then, by forcing $f''+cf'+f=0$, we find $fw''+(2f'+cf)w'=f^2g^2$.  We choose the linearly independent solution $f=\mathrm{e}^{\lambda z}$, where $\lambda=(-c+\sqrt{c^2-4})/2$, so that $w''+\sqrt{c^2-4}w'=\mathrm{e}^{\lambda z}g^2$.  The equations simplify by setting $w=w(s)$, $s=h(z)$.  The left-hand side of this differential equation reduces to a single term if $h''+\sqrt{c^2-4}h'=0$, which suggests we choose $h=\mathrm{e}^{-\sqrt{c^2-4}z}$.  Finally, with $c^2=25/6$ we end up with the second-order differential equation for $w$ to be $\mathrm{d}^2w/\mathrm{d}s^2=6w$ which, upon multiplying both sides by $\mathrm{d}w/\mathrm{d}s$, integrates directly to
\begin{equation}
\left(\frac{\mathrm{d}w}{\mathrm{d}s}\right)^2=4w^3-\omega,
\label{eq:odew}
\end{equation}
where $\omega$ is a constant.  The first-order ode (\ref{eq:odew}) is separable and is solved exactly in terms of the Weierstra$\ss$ p-function $\wp(z;0;\omega)$ \cite{Abramowitz1965}.  Rewriting the solution in terms of $U$ and $V$ gives
\begin{align}
U &= \mathrm{e}^{-2z/\sqrt{6}}\wp\left(\mathrm{e}^{-z/\sqrt{6}}-k;0;\omega\right), \label{eq:exact1}\\
V &=  -\frac{\sqrt{6}}{3}\mathrm{e}^{-2z/\sqrt{6}}\left(
\wp\left(\mathrm{e}^{-z/\sqrt{6}}-k;0;\omega\right)
+2\mathrm{e}^{-z/\sqrt{6}}\wp'\left(\mathrm{e}^{-z/\sqrt{6}}-k;0;\omega\right)
\right),
\label{eq:exact2}
\end{align}
where $\wp'$ is the derivative of $\wp$ \cite{Abramowitz1965}.  It is noteworthy that this exact solution is possible because (\ref{eq:2ndorderode}) has the Painlev\'{e} property for $c=\pm 5/\sqrt{6}$.  In other words, for these special values of $c$, the movable singularities of solutions to (\ref{eq:2ndorderode}) are poles~\cite{Clarkson2006}.

It is worth plotting the exact solution(s) (\ref{eq:exact1})-(\ref{eq:exact2}) both in the form $U=U(z)$ and in the phase-plane.  As the travelling wave solutions are invariant to translations in $z$, we fix each solution in the $z$-direction by setting $U=U_0$ at $z=0$.  For a given point in the phase plane, $(U,V)=(U_0,V_0)$, we determine the two constants $k$ and $\omega$ by solving the nonlinear algebraic system $U_0=\wp(1-k;0;\omega)$, $V_0=-(\sqrt{6}/3)\left(\wp(1-k;0;\omega)+2\wp'(1-k;0;\omega)
\right)$ numerically, for example with Newton's method.  The system can be sensitive and may require a close initial guess.

Fixing $c= 5/\sqrt{6}$ for the moment, travelling wave profiles $U(z)$ are shown in Figure~\ref{fig:figure1}(a), while corresponding trajectories in the phase-plane are shown in Figure~\ref{fig:figure1}(b).  Each of these curves could be drawn using the exact solution (\ref{eq:exact1})-(\ref{eq:exact2}) or just as easily be generated using numerical solutions to (\ref{eq:1storderodeU})-(\ref{eq:1storderodeV}).  As is well known, the phase-plane in Figure~\ref{fig:figure1}(b) is characterised by two fixed points, namely the saddle point at $(1,0)$ and the stable node at $(0,0)$.  All of the trajectories in Figure~\ref{fig:figure1}(b) enter the stable node $(0,0)$; a linearisation about $(0,0)$ demonstrates that $U\sim \mbox{const}\,\mathrm{e}^{-2z/\sqrt{6}}$ as $z\rightarrow\infty$.  As $z$ decreases, we see in both Figures~\ref{fig:figure1}(a) and (b) that, with one exception, each solution (thin coloured curves) blows up (with $U\rightarrow \pm\infty$) at a finite value of $z$.  This is because $\wp(\zeta;0,\omega)$ has a double pole at $\zeta=0$.  The exception is the heteroclinic orbit (thick black curve) that joins the two fixed points; this trajectory corresponds to the well-known exact solution (\ref{eq:exactsoln}), which notably satisfies the physically realistic boundary condition (\ref{eq:Urightarrow1}).  The simplification from (\ref{eq:exact1})-(\ref{eq:exact2}) to (\ref{eq:exactsoln}) in this case arises because this special case corresponds to taking the limit $k\rightarrow 0$, which is in effect {\em pushing} the singularity to $z=-\infty$.  Numerical solutions to (\ref{eq:FisherKPP})-(\ref{eq:FisherKPP_BC}) with appropriate initial conditions evolve to (\ref{eq:exactsoln}), as we demonstrate in section~\ref{sec:FSm}.

Now turning to $c= -5/\sqrt{6}$, we show results in Figure~\ref{fig:figure1}(c)-(d).  Here it is convenient to reflect our phase-plane about the $U$-axis, so the heteroclinic orbit just mentioned is in the upper-half plane.  The trajectories in the phase-plane are still given by (\ref{eq:exact1})-(\ref{eq:exact2}), except that we must make the changes $V\rightarrow -V$, $z\rightarrow -z$.  Five of the solutions shown in Figure~\ref{fig:figure1}(c)-(d) are the same as in Figure~\ref{fig:figure1}(a)-(b) (to enable a straightforward comparison across the figures, we have labelled these solutions $\textcircled{\footnotesize 1}$--$\textcircled{\footnotesize 5}$); these are trajectories that start at the stable node $(0,0)$,  except now we see that $U\sim \mbox{const}\,\mathrm{e}^{2z/\sqrt{6}}$ as $z\rightarrow -\infty$.  As we follow these trajectories for increasing $z$, we again note that they blow up at a finite value of $z$ (with $U\rightarrow -\infty$).  Other trajectories also blow up for finite $z$ (this time with $U\rightarrow \infty$), except for the separatrix (solid black curve) which eventually enters the saddle point $(1,0)$.

We now interpret the separatrix in Figure~\ref{fig:figure1}(d) in terms of an exact sharp-fronted travelling wave solution of (\ref{eq:FisherKPP}).  As explained in Ablowitz \& Zeppetella \cite{Ablowitz1979}, in order to extract this special case from (\ref{eq:exact1})-(\ref{eq:exact2}), we must choose the constants $k$ and $\omega$ such that $\wp(\zeta; 0,\omega)$ has one of its double poles at $\zeta=-k$. This is a numerical constraint which can be enforced by solving the system
\begin{equation}
U_0=\wp(1-k;0;\omega),
\quad
\wp'(k;0;\omega)/\wp(k;0;\omega)^2=0,
\label{eq:constraint}
\end{equation}
where $0<U_0<1$.  In Figure~\ref{fig:figure1}(c) we have chosen $U_0=0.2$, but of course this value is arbitrary and, in effect, corresponds to a translation in $z$. The key observation of this special solution is that it satisfies the physically realistic boundary condition (\ref{eq:Urightarrow1}) as $z\rightarrow -\infty$.  As we see in Figure~\ref{fig:figure1}(c), as $z$ increases, this solution for $U$ is very flat until it decreases sharply to $U=0$ at some finite value of $z=z^*$ where $V=V^*=-2.251\ldots$, and then continues to decrease as $z$ increases further.  While this solution satisfies (\ref{eq:Urightarrow1}), it is normally disregarded as it is negative for all $z>z_c$.  In the following section we show how this profile is a receding travelling wave solution for the Fisher-Stefan moving boundary problem (\ref{eq:StefanFisher1})-(\ref{eq:StefanFisher3}).


\section{Time-dependent solutions to the Fisher-Stefan model}\label{sec:FSm}

\begin{figure}
	\centering
	\includegraphics[width=1\textwidth]{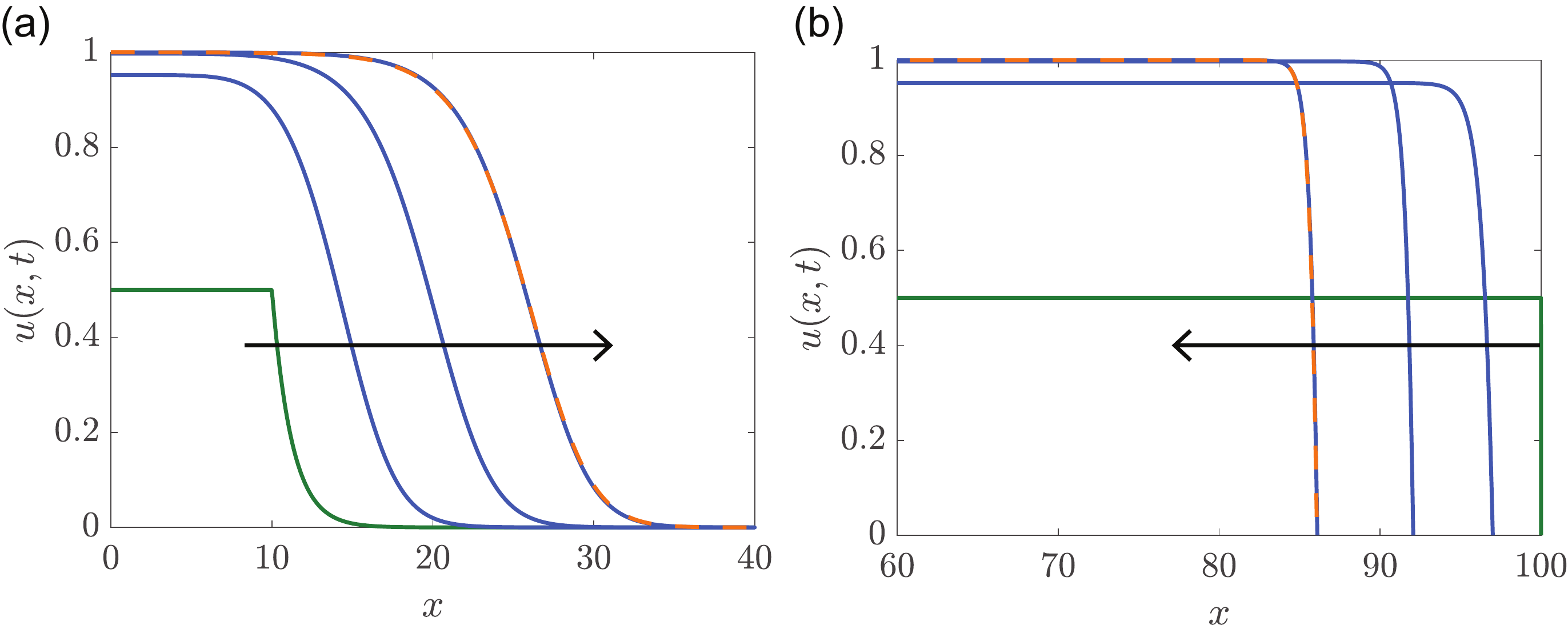}
	\caption{\textbf{Time-dependent PDE solutions}. (a) Numerical solution of (\ref{eq:FisherKPP})-(\ref{eq:FisherKPP_BC}) showing the initial condition (green),  intermediate-time solutions at $t=3$, $6$ and $9$ (blue).  The solution at $t=9$ is superimposed with (\ref{eq:exactsoln}) (dashed orange).  (b) Numerical solution of (\ref{eq:StefanFisher1})-(\ref{eq:StefanFisher3}) showing an initial condition (green), intermediate-time solutions at $t=3$, $6$ and $9$ (blue).  The solution at $t=9$ is superimposed with (\ref{eq:exact1})-(\ref{eq:exact2}) subject to (\ref{eq:constraint}) (dashed orange).}
	\label{fig:figure2}
\end{figure}

Figure \ref{fig:figure2}(a) shows time-dependent solutions of (\ref{eq:FisherKPP})-(\ref{eq:FisherKPP_BC}). Here, we see that a carefully-chosen initial condition with the appropriate decay at infinity evolves to a travelling wave solution that is visually indistinguishable from  (\ref{eq:exactsoln}).  To make this point we show the initial condition in green, together some intermediate-time solutions in blue.  The latest solution is superimposed with (\ref{eq:exactsoln}) in dashed orange, confirming that the time-dependent solutions converge to the exact solution reasonably rapidly.

Similar results in Figure \ref{fig:figure2}(b) show time-dependent solutions of (\ref{eq:StefanFisher1})-(\ref{eq:StefanFisher3}).  Here we have used a simple step function for an initial condition, which is shown in green, and have carefully chosen our parameter $\kappa$ to be $\kappa =  c/V^*=-0.906\ldots$.  Again we show three intermediate-time solutions in blue.  The superimposed exact solution (\ref{eq:exact1})-(\ref{eq:exact2}) with (\ref{eq:constraint}) for $c=-5/\sqrt{6}$ in orange compares extremely well with the numerical solution at $t=9$.

\section{Conclusion} \label{sec:Conclusion}

We present a new interpretation of an exact travelling wave solution of the Fisher-KPP model.  The Fisher-KPP model is one of the most well-studied reaction-diffusion equations with applications including wound healing~\cite{Johston2015,Sengers2007,Sherratt90,Vittadello2018} and ecological invasion~\cite{Skellam1951,Steele1998}.  For cell biology applications, the Fisher-KPP model is often used because cells are thought to move randomly, by diffusion, as well as proliferating logistically~\cite{Maini2004,Sherratt90}.  Experimental observations of moving cell fronts can be described by travelling wave solutions of the Fisher-KPP model~\cite{Maini2004}, or generalisations of the Fisher-KPP model~\cite{Warne2019}.  For the dimensional Fisher-KPP model with diffusivity $D$, proliferation rate $\lambda$ and carrying capacity density $K$,  the speed of the travelling wave solution is $c\geq 2\sqrt{\lambda D}$~\cite{Murray02}.  In reality, fronts of cells may move at a slower speed or even retreat~\cite{Elhachem2020b}.  One way to deal with this is to write down a more complicated model with more than one species~\cite{Painter2003} or with a different source term, like an Allee effect~\cite{Fadai2020}.  Even with a single species model that retains a logistic growth term, we can introduce nonlinear degenerate diffusion, leading to the Porous-Fisher model~\cite{Witelski1995}, which gives rise to travelling wave solutions with $c\geq\sqrt{\lambda D/2}$~\cite{Sanchez1994,Sherratt1996}.   An alternative to all of these modifications is to simply retain Fisher (\ref{eq:FisherKPP}) but include a moving boundary as in (\ref{eq:StefanFisher1})-(\ref{eq:StefanFisher3})~\cite{Du2010}.  This has appealing features, namely: a sharp moving front, which seems biologically reasonable; retains the classical logistic growth term with an easy to measure $\lambda$; allows for solutions of all speeds $c<2\sqrt{\lambda D}$, including negative speeds~\cite{Elhachem2019}.

Despite the apparent simplicity of the Fisher-KPP model, exact solutions are relatively elusive but of high interest since they provide important mathematical insight and can be used as benchmarks for testing numerical methods~\cite{Simpson2006}.  It is well-known that the travelling wave solution (\ref{eq:exactsoln}) can be written down exactly for a special wave speed $c=5/\sqrt{6}$~\cite{Ablowitz1979,Kaliappan1984}.  This special travelling wave is consistent with the usual view that travelling wave solutions of the nondimensional Fisher-KPP have positive speed  $c > 2$, whereas solutions with $c< 2$ are normally disregarded on the grounds of being unphysical~\cite{Murray02}.  In our work we take a different point of view and re-formulate the Fisher-KPP model with a moving boundary, often called the Fisher-Stefan model~\cite{Elhachem2019}.  The Fisher-Stefan model has several attractive features: (i) travelling wave solutions of the Fisher-Stefan model have a well-defined front without needing to introduce the complication degenerate nonlinear diffusion; (ii) the Fisher-Stefan model gives rise to travelling wave solutions with $\infty < c< 2$, which is more flexible that the usual Fisher-KPP and Porous-Fisher models since it can be used to model both invasion and retreat; and (iii)  the Fisher-Stefan model provides a simple physical interpretation for an exact solution with $c=-5/\sqrt{6}$.  This overlooked  solution can be expressed exactly using Weierstra$\ss$ elliptic functions and, as we show numerically, this solution is the long-time limit of our moving boundary problem (\ref{eq:StefanFisher1})-(\ref{eq:StefanFisher3}). \\

\newpage

\section{Appendix } \label{sec:Numericalmethods}
\subsection{Numerical method for the phase plane} \label{sec:Numericalmethods}
To construct the phase planes in the main document we solve the first order dynamical system
\begin{align}
\frac{\text{d}U}{\text{d} z} & = V, \label{eq:ODEdU}\\
\frac{\text{d}V}{\text{d} z} &= -cV - U(1-U), \label{eq:ODEdV}
\end{align}
numerically, using a two-stage Runge-Kutta method, also called Heun's method~\cite{Simpson2005}, with a constant step size, $\textrm{d}z = 1 \times 10^{-3}$.  This choice of discretisation leads to results that are visually independent of the numerical discretisation.  Our main interest is in examining phase plane trajectories that either enter or leave the saddle $(1,0)$ along the stable or unstable manifold, respectively.  Therefore, it is important that the initial condition we chose when solving Equations (\ref{eq:ODEdU})--(\ref{eq:ODEdV}) are on the appropriate stable or unstable manifold, and sufficiently close to $(1,0)$.  To choose this point we use the MATLAB \textit{eig} function~\cite{eig} to calculate the eigenvalues and eigenvectors when $c=\pm 5/\sqrt{6}$. The vector field associated with  Equations (\ref{eq:ODEdU})--(\ref{eq:ODEdV}) are plotted on the phase planes using the MATLAB \textit{quiver} function~\cite{quiver}.  MATLAB implementations of these algorithms are available on \href{https://github.com/maudelhachem/McCue2020}{GitHub}.

\subsection{Phase plane visualisation} \label{sec:PPVis}
\begin{figure}[h]
	\centering
	\includegraphics[width=1\textwidth]{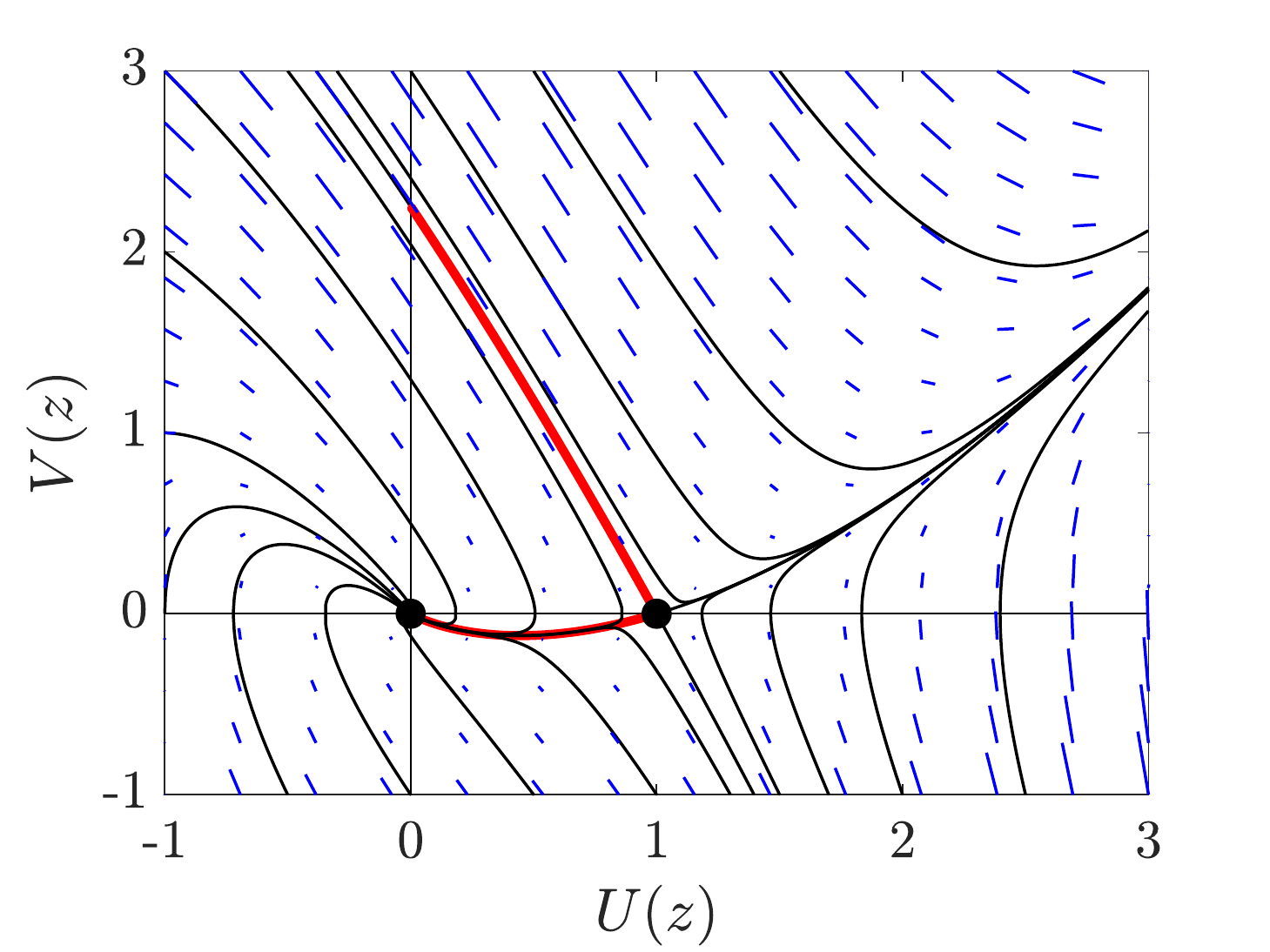}
	\caption{\textbf{Phase plane visualisation}. Phase plane showing the two equilibria (black discs).  The two trajectories corresponding to the travelling wave solutions (thick red lines) and a range of other trajectories (thin black lines) are shown.}
	\label{fig:figureS1}
\end{figure}

Results in Figure 1 (main document) show regions of the phase plane that are deliberately scaled so that we can easily see the details of the trajectories associated with $c = \pm 5 / \sqrt{6}$.  Since each phase plane in Figure 1 (main document) is shown on a different scale, it is easy to forget that these phase planes are closely related, with the only difference being the sign of $c$.  To make this point we show here, in Figure \ref{fig:figureS1}, the phase plane shown over a wider domain with $c = \pm 5 / \sqrt{6}$.  We show, in thick red lines, the trajectories associated with the travelling waves with $c = \pm 5 / \sqrt{6}$.  For completeness, we also show many other phase plane trajectories, in thin black lines.  These additional trajectories are not associated with travelling waves.  Note that the thick red trajectory in the third and fourth quadrant are associated with the invading travelling wave with $c=5/\sqrt{6}$, whereas the thick red trajectory in the first quadrant is associated with the receding travelling wave with $c=-5/\sqrt{6}$.  To plot these trajectories on the same phase plane, the trajectory associated with the receding travelling wave $c=-5/\sqrt{6}$ is obtained by integrating the dynamical system in the other direction, effectively by setting $\textrm{d}z = -1 \times 10^{-3}$.

\subsection{Numerical method for the Fisher--KPP model}\label{sec:FisherKPP}
We solve the Fisher--KPP equation
\begin{equation}\label{eq:FisherKPPn}
\frac{\partial u}{\partial t} =\frac{\partial^2 u}{\partial x^2} +  u(1-u),
\end{equation}
on $0 \le x \le L$, with $L$ chosen to be sufficiently large so that the time--dependent solutions of the partial differential equation (PDE) model have sufficient time to approach a travelling wave.  We discretise the domain with a uniform finite difference mesh, with spacing $\Delta x$, and we approximate the spatial derivatives in Equation (\ref{eq:FisherKPPn}) using a central finite difference approximation.  The resulting system of coupled ordinary differential equations are integrated through time using an implicit Euler approximation, giving rise to
\begin{equation}\label{eq:FD}
\frac{u_{i}^{j+1} - u_{i}^{j} }{\Delta t} = \left( \frac{u_{i-1}^{j+1}- 2 u_{i}^{j+1} + u_{i+1}^{j+1}}{\Delta x^2} \right) +  u_{i}^{j+1} (1-u_{i}^{j+1}),
\end{equation}
for $i = 2, \ldots, m-1$, where $m = L/\Delta x + 1$ is the total number of spatial nodes, and $j$ indexes time so that we approximate $u(x,t)$ by $u_{i}^{j}$, where $x=(i-1)\Delta x$ and $t=j\Delta t$.

For all numerical solutions of Equation (\ref{eq:FisherKPPn}) we enforce no--flux boundary conditions at $x=0$ and $x=L$, giving
\begin{align}
&u_{2}^{j+1}-u_{1}^{j+1} = 0,  \label{eq:FKpp_BC1} \\
&u_{m}^{j+1}- u_{m-1}^{j+1} = 0. \label{eq:FKpp_BC2}
\end{align}

In this work we wish to study the solution of Equation (\ref{eq:FisherKPPn}) with a travelling wave speed greater than the minimum speed, $c > 2$.  Therefore, care is taken to choose the initial condition to achieve this.  For the initial condition we set
\begin{equation}
\label{eq:ICFKPP}
u(x,0) =
\begin{cases}
\dfrac{1}{2} &\quad  \ 0 \le x < 10, \vspace{3mm} \\
\dfrac{\mathrm{e}^{-\dfrac{2}{\sqrt{6}}(x-10)}}{2} &\quad  \ 10 \le x < L, \\
\end{cases}
\end{equation}
where the exponential decay rate of the initial condition is carefully chosen to be $u(x,0) \sim \mathrm{e}^{-2x/\sqrt{6}}$, because this choice leads to travelling waves with $c=5/\sqrt{6}$~\cite{Murray02}.

To advance the discrete system from time $t$ to $t + \Delta t$ we solve the system of nonlinear algebraic system, Equations (\ref{eq:FD})--(\ref{eq:FKpp_BC2}), using Newton--Raphson iteration with convergence tolerance $\varepsilon$. The resulting systems of linear equations are solved efficiently using the Thomas algorithm. All numerical solutions of  Equation (\ref{eq:FisherKPPn}) are obtained by setting $\Delta x = 1 \times 10^{-4}$, $\Delta t = 1 \times 10^{-2}$, $L=60$ and $\varepsilon= 1 \times 10^{-8}$.  We find that these choices lead to results that are independent of the numerical discretisation.   Using the numerically--generated time--dependent solutions we  also estimate the travelling wave speed.  To estimate $c^*$ we specify a contour value, $u(x,t) = u^*$, and at the end of each time step, we use linear interpolation to find $x^*$ such that $u(x^*,t) = u^*$, and we then calculate$c^* = \left[x^*(t + \Delta t)-x^*(t)\right]/\Delta t$. MATLAB implementations of these algorithms are available on \href{https://github.com/maudelhachem/McCue2020}{GitHub}.

\subsection{Numerical method for the Fisher--Stefan model}\label{sec:FisherStefan}
We solve
\begin{equation}\label{eq:FisherKPPmovbound}
\frac{\partial u}{\partial t} =\frac{\partial^2 u}{\partial x^2} +  u(1-u),
\end{equation}
for $0 < x < s(t)$ numerically by using a boundary fixing transformation $\xi = x / s(t)$ so that we have
\begin{align}\label{eq:FisherKPPmovboundxi}
\frac{\partial u}{\partial t} = \frac{1}{s^2(t)} \frac{\partial^2 u}{\partial \xi^2}+\frac{\xi}{s(t)} \frac{\text{d}s(t)}{\text{d} t} \frac{\partial u}{\partial \xi} + u(1-u),
\end{align}
on the fixed domain, $0 < \xi < 1$.  Here $s(t)$ is the length of the domain that we will discuss later.  To close the problem we also transform the appropriate boundary conditions
\begin{align}
&\dfrac{\partial u}{\partial \xi} = 0 \quad \textrm{at} \quad \xi=0,  \label{eq:FS_BC1} \\
&u = 0 \quad \textrm{at} \quad \xi=1. \label{eq:FS_BC2}
\end{align}

Spatially discretising Equations (\ref{eq:FisherKPPmovboundxi})--(\ref{eq:FS_BC2}) with a uniform finite difference mesh, with spacing $\Delta \xi$, approximating the spatial derivatives using central differences and an implicit Euler approximation for the temporal derivative gives,
\begin{align}\label{eq:FDDinternalxi}
\dfrac{u_{i}^{j+1} - u_{i}^{j} }{\Delta t} &=  \dfrac{1}{(s^{j})^2} \left( \frac{u_{i-1}^{j+1} - 2 u_{i}^{j+1} + u_{i+1}^{j+1}}{\Delta \xi^2} \right)+ \dfrac{\xi}{s^{j}} \left(\dfrac{s^{j+1} - s^{j} }{\Delta t}\right) \left( \frac{u_{i+1}^{j+1} - u_{i-1}^{j+1}}{2 \Delta \xi} \right) +  u_{i}^{j+1}(1 - u_{i}^{j+1}) ,
\end{align}
for $i = 2, \ldots, m-1$, where $m = 1/\Delta \xi + 1$ is the total number of spatial nodes on the finite difference mesh, and the index $j$ represents the time index so that we approximate $u(\xi,t)$ by $u_{i}^{j}$, where $\xi~=~(i~-~1)~\Delta~\xi$ and $t=j\Delta t$.

Discretising the boundary conditions, Equations (\ref{eq:FS_BC1})--(\ref{eq:FS_BC2}), gives
\begin{align}
&u_{2}^{j+1}-u_{1}^{j+1} = 0,  \label{eq:FS_BC1a} \\
&u_{m}^{j+1} = 0. \label{eq:FS_BC2a}
\end{align}

The initial condition for the Fisher--Stefan problem is $u(x,0)=1$ on $0 \le x \le s(0)$. To advance the discrete system from time $t$ to $t + \Delta t$ we solve the system of nonlinear algebraic equations, Equations (\ref{eq:FDDinternalxi})--(\ref{eq:FS_BC2a}), using Newton--Raphson iteration with convergence tolerance $\varepsilon$. During each iteration of the Newton--Raphson algorithm we update $s(t)$ using the discretised Stefan condition
\begin{equation}
s^{j+1} = s^{j} - \dfrac{\kappa \Delta t}{s^j\Delta \xi} \left(\dfrac{u_{m-2}^{j+1}}{2}-2u_{m-1}^{j+1}+ \dfrac{3u_{m}^{j+1}}{2}\right),
\label{eq:supdatediscretise}
\end{equation}
where we set  $u_{m}^{j+1} = 0$ to be consistent with (\ref{eq:FS_BC2a}).

All results in this work are obtained by setting $\kappa=-0.906\ldots$, $\Delta \xi = 1 \times 10^{-5}$, $\Delta t = 1 \times 10^{-2}$, $\varepsilon = 1 \times 10^{-8}$ and $s(0)=100$. Again, we find these choices lead to results that are independent of the numerical discretisation, and we also use the time--dependent PDE solution to estimate $c^*$ by tracking the time evolution of the leading edge, $c^* = (s^{j+1} - s^{j})/\Delta t$. MATLAB implementations of these algorithms are available on \href{https://github.com/maudelhachem/McCue2020}{GitHub}.

\paragraph{Acknowledgements:}This work is supported by the Australian Research Council (DP200100177).

\end{document}